\documentclass[prb,aps,twocolumn,showpacs]{revtex4-1}
\usepackage[OT4]{fontenc}

\usepackage{graphicx}

\usepackage{amsmath,bm,amsfonts}
\newcommand{\rr}{\bm{r}}
\newcommand{\dd}{\bm{\hat{d}}}
\newcommand{\bd}{\bm{d}}

\newcommand{\lon}{\mathrm{l}}
\newcommand{\tra}{\mathrm{t}}
\newcommand{\kl}{k_{\mathrm{l}}}
\newcommand{\kt}{k_{\mathrm{t}}}

\DeclareMathOperator{\im}{Im}
\begin{document}
\author{A. Musia{\l}}
\author{P. Kaczmarkiewicz}
\author{G. S\k{e}k}
\author{P. Podemski}
\author{P. Machnikowski}
\author{J. Misiewicz}
 \affiliation{Institute of Physics, Wroc{\l}aw University of
Technology, 50-370 Wroc{\l}aw, Poland}
\author{S. Hein}
\author{S. H\"ofling}
\author{A. Forchel}
\affiliation{Technische Physik, Universit\"at W\"urzburg and Wilhelm
  Conrad R\"ontgen-Center for Complex Material Systems, Am Hubland,
  D-97074 W\"urzburg, Germany} 

\title{Carrier trapping and luminescence polarization in quantum dashes}

\begin{abstract}
We study experimentally and theoretically polarization-dependent
luminescence from an ensemble of quantum-dot-like nanostructures with
a very large in-plane shape anisotropy (quantum dashes). We show that the
measured degree of linear polarization of the emitted light increases
with the excitation power and changes with temperature in a
non-trivial way, depending on the excitation conditions. Using an
approximate model based on the 
$\bm{k}\cdot\bm{p}$ theory, we are able to relate this degree of
polarization to the amount of light hole admixture in the exciton
states which, in turn, depends on the symmetry of the envelope wave
function. Agreement between the measured properties and theory is
reached under assumption that the ground exciton state in a quantum
dash is trapped in a confinement fluctuation within the structure
and thus localized in a much smaller volume of much lower
asymmetry than the entire nanostructure. 
\end{abstract}


\maketitle

\section{Introduction}

Quantum dashes (QDashes) are epitaxially grown nanostructures strongly
elongated in one of the in-plane directions.
They can be spontaneously and
preferentially formed in a molecular beam epitaxy process of
self-assembled growth of, e.g., InAs on InP
substrate (Ref.~\onlinecite{reithmaier07} and 
references therein). 
The interest in these structures is partly motivated by their
applications in telecommunication lasers due to their higher surface
density, favorable emission wavelength, wide spectral tunability,
broad gain, and high speed modulation
\cite{reithmaier07,djie08,sauerwald05,hein09}.
The existing theoretical predictions on some
quantum dash properties include calculation of their electronic states
as well as the transition dipole moments and the resulting optical
spectra \cite{miska04,wei05,planelles09}. Experimentally,
the optical properties of InAs/InP 
QDashes have been investigated from the point of view of both the
ensemble \cite{rudno06} and the single object properties
\cite{Mensing03,sek09}. So far, however, the details of the spatial
character of the confining potential and polarization properties of
these structures have not been analyzed.

\begin{figure}[!tb]
\begin{center}
  \includegraphics[width=60mm]{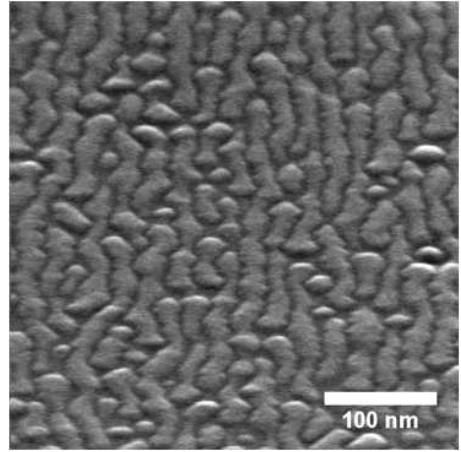}
\end{center}
\caption{\label{fig:SEM}Scanning electron microscopy image of a
QDash  sample before overgrowth.} 
\end{figure}

A scanning electron microscopy (SEM) image of a QDash sample, shown in 
Fig.~\ref{fig:SEM}, reveals 
shape irregularities of these structures in the form of sections with locally
increased thickness as well as zig-zag bends. While the shape
irregularity is a generic feature of QDash structures, the exact
morphology varies from sample to sample (depending on the details of
the growth procedure) and either thickness fluctuations or bends can
be the dominating irregularity feature \cite{reithmaier07,ooi08}. Both
the thick 
sections and the zig-zag corners \cite{intonti01}, as well as the possible
composition inhomogeneities, can act as additional
trapping centers within the confinement volume of the elongated
nanostructure. Such an effect of confining in an effective potential
on length scales smaller than the size of the entire nanostructure was
observed in  V-shaped
quantum wires\cite{guillet03}. In quantum dashes,
the localized character of emission at 
low temperatures in low excitation regime is suggested by
experimental observations on single QDashes, where some features
typical for the strong confinement regime have been
unexpectedly revealed \cite{sek09} in spite of the large volume of
these structures. First of all, the obtained biexciton
binding energy is approximately 0.5 meV, which is a value
characteristic for very small or very large quantum objects
\cite{narvaez05}. The latter possibility can, however, be excluded by
noting that the estimated exciton to biexciton
lifetime ratio is close to 2, which is a clear fingerprint of a rather strong
confinement regime \cite{narvaez06,wimmer06}. 

The presence of 
an additional confinement within a larger structure can be seen in the
temperature dependence of the emission lifetime \cite{guillet03}
or in high resolution photoluminescence mapping by using near field
optical spectroscopy \cite{intonti01}. However, these methods 
require
time-resolved spectroscopy
techniques that suffer from the lack of high sensitivity detectors in
the 1.55~$\mu$m range (necessary for the investigation of dashes on the InP
substrate) or an AFM-based near field spectroscopy apparatus. Therefore,
a different approach for identifying such a two-stage
character of the localizing potential in QDash structures is highly desired. 
Such a method can be based on the fact that the anisotropy of the
confinement is reflected in the 
polarization properties of the luminescence
\cite{jo10,mensing06,ridha07,podemski08,yu99,tonin11}.
This relation can be traced back to subband mixing effects:
Oppositely circularly polarized contributions to the optical emission
originating 
from heavy and light hole transitions interfere, leading to
elliptically polarized emission and to the appearance of the
preferential orientation of the linear polarization
\cite{koudinov04,leger07}. 
A detailed analysis of the carrier states in an elongated
structure, based on the multi-band $\bm{k}\cdot\bm{p}$ theory
\cite{andrzejewski10,saito08} or 
empirical tight-binding approach \cite{sheng08} indeed reproduces the
polarization properties of the observed emission from anisotropic
structures.
The presumed localizing minima in a QDash 
can be expected to be significantly more
isotropic than the whole QDash itself. 
Our idea is, therefore, to use the degree of linear polarization
of the continuous wave (cw) luminescence of the 
system as the indicator of the localization of
carriers. 

In this work, we show that indeed the emission from the states confined in
the trapping centers has much more isotropic polarization properties
than the radiation originating from states delocalized over the whole
QDash volume due to a considerably different degree of anisotropy of
their  wave functions. For experimental confirmation, we investigate
the linear-polarization-resolved photoluminescence (PL) from InAs/InP
QDashes for different carrier distributions obtained by changing the
temperature and excitation power showing that the polarization study
of the QDash luminescence can yield information on the spatial
character of the confining potential and reveal the presence of an
additional carrier (exciton) trapping within a QDash. 

The paper is organized as follows. In Sec.~\ref{sec:exp}, we
describe the system and the experimental
setup and present preliminary PL results motivating the following
study. Next, in 
Sec.~\ref{sec:theory-mixing}, we present a 
general theory relating the polarization properties to the hole
subband mixing, followed by a description of our model of a QDash
(Sec.~\ref{sec:theory-model}). 
Sec.~\ref{sec:results} presents the results of our
polarization-dependent PL measurements and theoretical modeling. The
final Sec.~\ref{sec:concl} concludes the paper.

\section{Sample and experimental setup}
\label{sec:exp}

The experiment was performed on an ensemble of self-assembled InAs
quantum dashes 
grown epitaxially on a $(001)$ InP substrate with 3.4\% lattice
mismatch in the molecular beam epitaxy technique. The QDash layer
is surrounded by additional In$_{0.53}$Ga$_{0.23}$Al$_{0.24}$As quaternary
barriers, lattice matched to InP. The resulting nanostructures are
significantly elongated in one of the in-plane directions,
preferentially in $[1\overline{1}0]$\cite{reithmaier07}. 
In Fig.~\ref{fig:SEM}, we present an example of a scanning electron
microscopy 
image of a QDash sample. It should be kept in mind, however, that
exact morphology of these structures may vary depending on the growth
conditions (see Refs.~\onlinecite{reithmaier07,ooi08}). The lateral
dimensions, estimated for a sample as in Fig.~\ref{fig:SEM}, are
approximately several to tens of 
nanometers in width and even hundreds of nanometers in length,
confirming the expected significant shape asymmetry (lateral aspect ratio
above 4). Their height is typically on the order of a few
nanometers. This geometry, formed during the self-assembled growth, is
a result of a diffusion coefficient anisotropy due to the surface
reconstruction. Further growth details can be found elsewhere
\cite{reithmaier07,sauerwald05}. For the experimental studies, dashes
obtained by deposition 
of approximately 1~nm of InAs have been chosen, yielding typically the
room temperature emission in the range of 1.55~$\mu$m. The surface
density of these structures in the sample is rather high and exceeds
$10^{11}$~cm$^{-2}$. 

The linear-polarization-resolved photoluminescence measurements were
performed in a standard photoluminescence setup. The structures were
non-resonantly excited by a continuous wave semiconductor laser at the
wavelength of 660~nm. 
The excitation powers ranging from 
$P_{0}=100$~nW to $10^{4}P_{0}=1$~mW (determined outside the cryostat)
were 
used. The beam was focused onto a 
spot of approximately 0.01~mm$^{2}$ on the sample surface.
The emitted light was dispersed in a 0.5~m focal
length monochromator and the signal was detected by a liquid nitrogen
cooled InGaAs CCD linear detector. The 
PL spectra were measured for the two orthogonal
polarization directions: along and perpendicular to the quantum dash
elongation axis. From these measurements, the degree of linear
polarization (DOP), defined as 
\begin{equation}\label{DOP}
\Pi=
\frac{I_{\mathrm{l}}-I_{\mathrm{t}}}{I_{\mathrm{l}}+I_{\mathrm{t}}},
\end{equation}
was determined, where $I_{\mathrm{l}}$ and $I_{\mathrm{t}}$ are the
PL intensities for the linear 
polarization parallel and perpendicular to the dash elongation axis,
respectively. In this work, we choose to study the DOP calculated
using the intensities at the maximum
of the ensemble luminescence peak. Using intensities integrated over
the whole spectral range leads to nearly identical results.

\begin{figure}[!tb]
\begin{center}
\includegraphics[width=85mm]{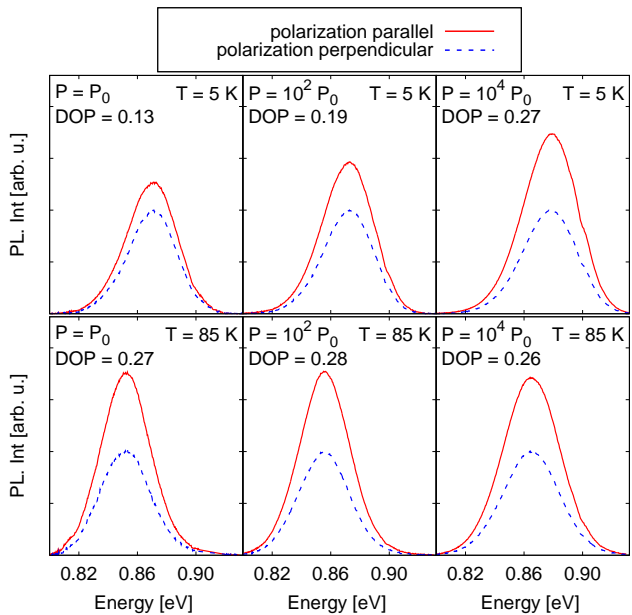}
\end{center}
\caption{\label{fig:eksp-PL}(Color online) Polarization-resolved photoluminescence
  spectra measured on the ensemble of InAs/InP quantum dashes  for
  different excitation powers (determined outside  the cryostat) and temperatures. 
The obtained values of the DOP at the peak
  of the PL spectrum are given in the figures. The spectra are
  normalized in such a way that the maximum of the signal for the
  perpendicular polarization is the same for all three powers at a
  given temperature.} 
\end{figure}

A set of photoluminescence spectra, measured at two
temperatures for various excitation powers are presented in
Fig.~\ref{fig:eksp-PL}. For each temperature and
excitation power, two emission 
spectra for the two orthogonal polarization directions ($[110]$ and
$[1\overline{1}0]$, i.e., perpendicular and parallel to the quantum dash
longer in-plane dimension, respectively) were measured. 
The first set of the spectra (upper panels in Fig.~\ref{fig:eksp-PL}) have
been  obtained at low temperature for three different excitation
powers and normalized, while  
the second set (lower panels) corresponds to a higher temperature. 
It can be seen that in all the cases 
the polarization along quantum dashes dominates. However, at low
temperatures, the ratio
between the intensities of the linear polarization components
increases with the excitation power. As a figure of merit we use the
DOP defined in Eq.~\eqref{DOP}.
As can be seen in Fig.~\ref{fig:eksp-PL}, the obtained
DOP differs by more than a factor of two 
between the lowest and the highest excitation power used. 

Before we proceed to study the polarization-related features in detail
let us note that, based on the PL spectra shown in Fig.~\ref{fig:eksp-PL},
each QDash appears to behave as an isolated system, that is, coupling
between the dashes is negligible.
Indeed, if the inter-QDash coupling were strong enough
to induce considerable carrier transfer on the time scales of the
exciton life time the emission at low excitation powers would be
dominated by those dashes in which the exciton energy is the lowest,
while higher energy QDashes would contribute at higher excitation due
to Pauli blocking (in analogy to the appearance of emission from
higher shells in a single nanostructure). This would lead to considerable
power-dependent broadening of 
the ensemble photoluminescence which is not present in the
PL spectra (less then 10\% increase of the width is observed between
the lowest and highest powers used in Fig.~\ref{fig:eksp-PL}). 

In addition, only a slight blue shift of the ensemble PL feature is
seen in  Fig.~\ref{fig:eksp-PL} for increasing powers (about $7$~meV
between the lowest and the  highest powers used).
This allows us to
exclude heating effects even at the highest powers used, as this would
be accompanied with a red shift following the band gap reduction
according to the Varshni law.
An additional argument against heating effects follows from micro-PL
measurements on the same sample \cite{sek09}, where
much more focused beams, hence much higher power densities are
applied. In spite of this, the single QDash lines visible in those
experiments show no red shift that would be a signature of heating.

\section{Theory}
\label{sec:theory}

In this section, we first develop a general description of
luminescence polarization due to a confinement anisotropy in a
nanostructure (Sec.~\ref{sec:theory-mixing}) and later introduce a
specific model of electron, hole and exciton wave functions in a QDash
(Sec.~\ref{sec:theory-model}). Although the general theory of
Sec.~\ref{sec:theory-mixing} can be reduced to a simple formula that
may be able to qualitatively capture the essential properties of the
system one needs a more accurate characterization of the actual wave
functions, as proposed in Sec.~\ref{sec:theory-model}, in order to
quantitatively account for the experimentally observed polarization.

\subsection{Polarization orientation and hole 
subband mixing}  
\label{sec:theory-mixing}

For the calculation of Coulomb-correlated optically excited states of
a semiconductor system it is
most common and convenient to use the \textit{hole picture}, that is, to
describe the many-particle 
states of the almost fully occupied valence band in terms of the few
unoccupied states.
Single particle (electron and hole) states confined in a nanostructure
can then be approximately described in the single-band envelope
function approximation by the envelope wave functions
$\psi_{\mathrm{c(v)}i\lambda}(\rr)$, where $\lambda$ denotes
the subband, $i$ labels different eigenfunctions of the
confinement potential, and c(v) refers to the conduction (valence)
band. Note that in this notation $\psi_{\mathrm{v}i\lambda}$ is the
hole wave function (rather than an electron wave function for a valence bad
state). 
On the other hand, for multiple-band calculations, like
the $\bm{k}\cdot\bm{p}$-based perturbation theory to be presented
below, as well as for a formal discussion of inter-band transitions,
it is by far more convenient to use the \textit{electron picture}, in
which the states of electrons in the valence band are directly
represented by their wave functions. The hole wave function and the
valence band electron wave function are simply related by complex
conjugation. Thus, in the
electron picture, the envelope wave functions for the valence
band electron states are $\psi^{*}_{\mathrm{v}i\lambda}(\rr)$.
The corresponding creation and annihilation
operators (in the electron picture) will be denoted by 
$ a_{\mathrm{c(v)}i\lambda}^\dag,a_{\mathrm{c(v)}i\lambda}$. 

If the Coulomb interaction and hole subband mixing is included the
energy eigenstates of a confined exciton can be written in the general form 
\begin{equation}\label{wzor:exciton}
| X^{(\beta)} \rangle = \sum_{\substack{ij\\ \lambda \lambda'}} 
c_{(i\lambda)(j\lambda')}^{(\beta)}
a_{\mathrm{v}i\lambda}a^\dag_{\mathrm{c}j\lambda'} 
| 0 \rangle,
\end{equation}
where $\lambda$ and $\lambda'$ run through the valence and conduction
subbands, respectively,  $\beta$ indicates different exciton states,
$c_{(i\lambda)(j\lambda')}^{(\beta)}$ are the coefficients 
for the expansion in single-subband single-particle states, and
$|0\rangle$ is the ground state of the crystal.

The polarization of the light emitted in the recombination process
from a conduction band state to a valence band state is determined by
the interband matrix element of the positive frequency part of the
dipole moment operator $\dd$ (Ref.~\onlinecite{haug04}),
\begin{displaymath}
\dd = \bd_{\lambda\lambda'} 
\int d^3r \psi_{\mathrm{v}i\lambda}(\rr)\psi_{\mathrm{c}j\lambda'}(\rr)
a_{\mathrm{v}i\lambda}^\dag a_{\mathrm{c}j\lambda'},
\end{displaymath}
where $\bd_{\lambda\lambda'}$ is the interband matrix element between
the states at the $\Gamma$ point of the Brillouin zone in the bulk
material (note that, according to our notation,
$\psi_{\mathrm{v}i\lambda}(\rr)$ is the complex-conjugated valence
band wave function).  
The dipole moment for the optical transition is therefore
\begin{equation}\label{wzor:dipmom}
\langle0|\dd|X^{(\beta)}\rangle = 
- \sum_{\lambda\lambda'} \bd_{\lambda\lambda'} 
\alpha_{\lambda\lambda'}^{(\beta)},
\end{equation}
where
\begin{equation}\label{wzor:defalpha}
\alpha_{\lambda\lambda'}^{(\beta)} = 
\sum_{ij} c_{(i\lambda)(j\lambda')}^{(\beta)} 
\int d^3r \psi_{\mathrm{v}i\lambda}({\bm{r}}) 
\psi_{\mathrm{c}j\lambda'}({\bm{r}}).
\end{equation}

Let us assume that the initial state is spin-up ($\lambda'=+1/2$). The
only non-vanishing in-plane components of the dipole moment are to the
$+3/2$ heavy hole (hh) band and to the $-1/2$ light hole (lh) band,
with the corresponding bulk matrix elements along the crystallographic
axes\cite{haug04,axt94b}
\begin{equation}\label{wzor:dipole}
\bd_{3/2,1/2} = \frac{d_0}{\sqrt{2}}\left (\begin{array}{c}
-1\\
i\end{array} \right ), \quad \bd_{-1/2,1/2} = 
\frac{d_0}{\sqrt{6}}\left (\begin{array}{c}
1\\
i\end{array} \right ).\nonumber
\end{equation}
Note that the magnitude of the dipole matrix elements, $d_{0}$, is
irrelevant for the present study since the DOP is a relative quantity,
as follows from Eq.~\eqref{DOP}.

Since the studied system consists of QDashes elongated in the $[1 \bar
1 0]$ direction we define unit vectors parallel and transverse to this
direction, 
\begin{equation}\label{wzor:kierunki}
\hat e_{l} = \frac{\hat e_x - \hat e_y}{\sqrt{2}}, \quad 
\hat e_{t} = \frac{\hat e_x + \hat e_y}{\sqrt{2}},
\end{equation}
where $\hat{e}_{x},\hat{e}_{y}$ are the unit vectors along the
crystallographic axes.
From Eqs. (\ref{wzor:exciton}), (\ref{wzor:dipmom}) and
(\ref{wzor:kierunki}), the corresponding components of the interband
dipole moment are 
\begin{eqnarray}\label{wzor:dl}
d_l^{(\beta)} &=& \hat e_{l} \cdot \langle 0 | \dd | X^{(\beta)} \rangle \nonumber \\
&=& -d_0 \frac{i+1}{2}\alpha_{3/2,1/2}^{(\beta)} 
+ d_0 \frac{1-i}{2\sqrt{3}}\alpha_{-1/2,1/2}^{(\beta)}
\end{eqnarray}
and
\begin{eqnarray}\label{wzor:dt}
d_t^{(\beta)} &=& \hat e_{t} \cdot \langle 0 | \dd | X^{(\beta)} \rangle \nonumber \\
&=& d_0 \frac{i-1}{2}\alpha_{3/2,1/2}^{(\beta)} 
+ d_0 \frac{1+i}{2\sqrt{3}}\alpha_{-1/2,1/2}^{(\beta)}.
\end{eqnarray}

If the average occupation of an exciton state with energy $E$ is $n(E)$
then the intensities of light emitted from the state $\beta$, polarized
parallel and perpendicular to the direction of the QDash elongation are
proportional to  $I_{\mathrm{l}}^{(\beta)}\sim n(E_{\beta})|d_\mathrm{l}^{(\beta)}|^2$ and
$I_{\mathrm{t}}^{(\beta)}\sim n(E_{\beta})|d_\mathrm{t}^{(\beta)}|^2$,
respectively.

In order to provide a description relevant to an inhomogeneously
broadened ensemble we consider a set of QDashes with variable
parameters that we formally jointly denote by a single symbol
$\eta$, representing the full set of relevant parameters. The random
distribution of the QDash parameters is represented by the
distribution function  $f(\eta)$. At this point, the number and the
physical nature of the
variable parameters as well as the distribution function can
be arbitrary but a simple choice will be proposed in
Sec.~\ref{sec:theory-model} in order to perform numerical
calculations. 

For the PL response recorded from an inhomogeneous QDash
ensemble at a single emission energy $E$,
the PL intensities at the two linear polarizations are  then
\begin{displaymath}
I_{\mathrm{l/t}}(E)=
\int d\eta f(\eta)\sum_{\beta}n(E_{\eta\beta})
\left| d_{\mathrm{l/t}}^{(\eta\beta)} \right|^{2}\delta(E_{\eta\beta}-E),
\end{displaymath}
where $E_{\eta\beta}$ is the energy of the state
$|X^{(\beta)}\rangle$ and $d_{\mathrm{l/t}}^{(\eta\beta)}$ are the
corresponding components of the dipole moment in a QDash characterized
by the parameters $\eta$.
According to Eq.~\eqref{DOP} and using Eqs.~\eqref{wzor:dl}
and~\eqref{wzor:dt}, the ensemble DOP at the emission energy $E$ is
then
\begin{widetext}
\begin{equation}
\label{wzor:DOP}
\Pi(E) = \
-\frac{2}{\sqrt{3}} \frac{\int d\eta f(\eta)
\sum_{\beta} n_{\eta}(E_{\eta\beta})\delta(E_{\eta\beta}-E) 
\im \left [ \alpha_{3/2,1/2}^{(\eta\beta)*}
\alpha_{-1/2,1/2}^{(\eta\beta)}\right
]}{ \int d\eta f(\eta)
\sum_\beta n_{\eta}(E_{\eta\beta}) \delta(E_{\eta\beta}-E)
\left[ |\alpha_{3/2,1/2}^{(\eta\beta)}|^2
+\frac{1}{3}|\alpha_{-1/2,1/2}^{(\eta\beta)}|^2 \right ]},
\end{equation}
\end{widetext}
where $\alpha_{\lambda\lambda'}^{(\eta\beta)}$ is the
oscillator strength parameter, calculated according to
Eq.~\eqref{wzor:defalpha} for a QDash with parameters $\eta$
and $n_{\eta}(E_{\eta\beta})$ are the average occupations of the energy
levels in such a QDash (occupation redistribution is assumed to be
independent for each QDash, that is, no redistribution between the
dashes takes place).
As can be seen from the above equations, mixing between light and
heavy hole states leads to preferential polarization of the emitted
light, which depends on the relative phase of the two
contributions\cite{koudinov04,leger07}.  

In our theoretical modeling, we assume that the lowest hole state
consists mostly of a heavy hole component with some admixture from
light holes, as is typical for self assembled structures, where the
strain is compressive. This is justified in many systems since the light hole
states are shifted in energy with respect to the heavy hole states due
to confinement and strain, while the inter-subband coupling elements are
relatively small. In order to derive the light hole contribution one
uses a perturbation theory based on the $\bm{k}\cdot\bm{p}$ (Kane)
Hamiltonian \cite{winkler03}. As can be seen from the structure of the Kane
Hamiltonian, the heavy hole $+3/2$ state is coupled in the leading order to
both light hole states. However, the exciton state involving  a spin
$+1/2$ electron and a spin
$+1/2$ hole is dark and does not contribute to the optical properties
in the leading order. The calculations of the degree of polarization
[Eq.~(\ref{wzor:DOP})] have been performed with an assumption that
coupling to the $+1/2$ light hole states can be neglected. 

The part of the Kane Hamiltonian relevant to the light hole admixture
is 
\begin{displaymath}
V=\sum_{kl}\int
d^{3}r\psi_{\mathrm{v}l,3/2}(\bm{r})R\psi^{*}_{\mathrm{v}k,-1/2}(\rr) 
a^{\dag}_{\mathrm{v}l,3/2}a_{\mathrm{v}k,-1/2},
\end{displaymath}
where $R=R_{\mathrm{k}}+R_{\mathrm{s}}$ is the matrix element
of the Kane Hamiltonian coupling the spin 3/2 heavy hole subband with
the $-1/2$ light hole subband \cite{winkler03,andrzejewski10}. The
kinetic contribution to the inter-subband coupling is
\begin{displaymath}
R_{\mathrm{k}}  =  \frac{\sqrt{3}\hbar^2}{2 m_0}[\gamma_2(k_x^2-k_y^2)
-2i\gamma_3k_xk_y]
\end{displaymath}
where $k_j=-i \partial/\partial x_j$, $m_0$ is free electron mass and
$\gamma_j$ are the Luttinger parameters.
The strain-induced contribution is
\begin{displaymath}
R_{\mathrm{s}}  = 
- \frac{\sqrt{3}}{2}b_{\mathrm{v}}\left( \epsilon_{xx} -\epsilon_{yy}\right) 
+id_{\mathrm{v}}\epsilon_{xy},
\end{displaymath}
where $b_{\mathrm{v}}$ and $d_{\mathrm{v}}$ are valence
band deformation potentials and $\epsilon_{ij}$ are the strain tensor
components. 
Note that we
use the standard definition of the basis 
functions \cite{winkler03}, consistent with the general theory of the
angular momentum \cite{sakurai94}, which differs from that used in
many papers employing the $\bm{k}\cdot\bm{p}$ theory
\cite{pryor98,andrzejewski10}.

Assuming the zeroth-order exciton
state to be of purely heavy hole type with, say, a spin-up electron,
\begin{displaymath}
| X^{(\beta,0)} \rangle = \sum_{ij} 
c_{(i,3/2)(j,1/2)}^{(\beta)} a_{\mathrm{v}i,3/2}a^\dag_{\mathrm{c}j,1/2} | 0 \rangle,
\end{displaymath}
the first order correction due to the inter-subband coupling 
is approximated as
\begin{displaymath}
| X^{(\beta,1)} \rangle  = 
\sum_{kj} c_{(k,-1/2)(j,1/2)}^{(\beta)}
 a_{\mathrm{v}k,-1/2}a^\dag_{\mathrm{c}j,1/2} | 0 \rangle,
\end{displaymath}
with
\begin{eqnarray*}
\lefteqn{c_{(k,-1/2)(j,1/2)}^{(\beta)}=}\\
&&-\frac{1}{\Delta E_{\mathrm{lh}}}\sum_{i}
c_{(i,3/2)(j,1/2)}^{(\beta)}
\int d^{3}r \psi_{\mathrm{v}i,3/2}(\bm{r})
R\psi^{*}_{\mathrm{v}k,-1/2}(\rr),
\end{eqnarray*}
where $\Delta E_{\mathrm{lh}}$ is the average energy separation
between heavy and light holes.
Substituting this to Eq.~\eqref{wzor:defalpha} one finds
$\alpha_{-1/2,1/2}^{(\beta)}=
\alpha_{-1/2,1/2}^{(\beta,\mathrm{k})}+\alpha_{-1/2,1/2}^{(\beta,\mathrm{s})}$,
where $\alpha_{-1/2,1/2}^{(\beta,\mathrm{k})}$ and
$\alpha_{-1/2,1/2}^{(\beta,\mathrm{s})}$ are the kinetic and
strain-induced contributions, respectively, given by
\begin{eqnarray}\label{wzor:defalpha2}
\alpha_{-1/2,1/2}^{(\beta,x)} &=&  -\frac{1}{\Delta E_{\mathrm{lh}}}
\sum_{ijk} c_{(i,3/2)(j,1/2)}^{(\beta)}  \nonumber \\
& &\times \int d^3r \psi_{\mathrm{v}i,3/2}({\bm{r}}) R_{x} 
\psi^{*}_{\mathrm{v}k,-1/2}({\bm{r}})
\nonumber \\
&&\times \int d^3r \psi_{\mathrm{v}k,-1/2}({\bm{r}}) 
\psi_{\mathrm{c}j,1/2}({\bm{r}}) \nonumber \\
& = & -\frac{1}{\Delta E_{\mathrm{lh}}}
\sum_{ij} c_{(i,3/2)(j,1/2)}^{(\beta)}  \nonumber \\
& &\times \int d^3r \psi_{\mathrm{v}i,3/2}({\bm{r}}) R_{x} 
\psi_{\mathrm{c}j,1/2}({\bm{r}}),
\end{eqnarray}
for $x=\mathrm{k,s}$,
where we used the completeness relation for the wave functions
$\psi_{\mathrm{v}k,-1/2}(\rr)$. 

As the wave functions can be chosen real, $\alpha_{3/2,1/2}^{(\beta)}$
is real and only the imaginary part of $\alpha_{-1/2,1/2}^{(\beta)}$,
related to the imaginary (anti-hermitian) parts of $R_{\mathrm{k}}$
and $R_{\mathrm{s}}$,
contributes to Eq.~\eqref{wzor:DOP}.
In the
leading order in the subband-mixing term one then finds the DOP in
the form $\Pi=\Pi_{\mathrm{k}}+\Pi_{\mathrm{s}}$, where we split the effect
into the kinetic and strain-induced contributions defined by
Eq.~\eqref{wzor:DOP} with $\alpha^{(\eta\beta)}_{-1/2,1/2}$ replaced
in the numerator by its kinetic or strain-related parts
$\alpha^{(\eta\beta,x)}_{-1/2,1/2}$, $x=\mathrm{k,s}$, respectively,
and neglected in the denominator. 

The imaginary
(anti-hermitian) part of $R_{\mathrm{k}}$ can be written in the
coordinate frame 
related to the QDash elongation as
\begin{equation}\label{wzor:kpim}
\im R_{\mathrm{k}} = - \frac{\sqrt{3}\hbar^2}{2 m_0} 
\gamma_3(k_{\mathrm{l}}^2-k_{\mathrm{t}}^2),
\end{equation}
where $k_{\mathrm{l}} = (k_x-k_y)/\sqrt{2}$ and 
$k_{\mathrm{t}} = (k_x+k_y)/\sqrt{2}$.
From Eqs.~\eqref{wzor:DOP} and \eqref{wzor:defalpha2} it is then clear
that the DOP is determined by the matrix elements of 
$k_{\mathrm{l}}^{2}$ and $k_{\mathrm{t}}^{2}$ between the states
involved in the optical transition, hence by the symmetry of the wave
function. 

In order to estimate the strain-related part one would need to know
the exact strain distribution in the nanostructure, which is beyond
the scope of the present paper. Nonetheless, if the localization is
due to relatively small and smooth shape fluctuations then the strain
field can be expected to vary only very weakly over the volume of the
QDash (note that the situation can be completely different if the
trapping takes place in corners or bends on the QDash where the strain
distribution is likely to be much different from that in the straight
segments). Here, we will adopt the assumption of a nearly constant
strain field. Then, comparing Eq.~\eqref{wzor:defalpha2} for
$R_{\mathrm{s}}=\mathrm{const}$ with Eq.~\eqref{wzor:defalpha} one
immediately gets 
\begin{displaymath}
\alpha_{-1/2,1/2}^{(\beta,\mathrm{s})}=
-\frac{R_{\mathrm{s}}}{\Delta E_{lh}}\alpha_{3/2,1/2}^{(\beta)}.
\end{displaymath}
The strain-induced contribution to the DOP in this case is
\begin{displaymath}
\Pi_{\mathrm{s}}=
\frac{2}{\sqrt{3}}\frac{\im R_{\mathrm{s}}}{\Delta E_{lh}}
\end{displaymath}
and is the same for every state.

While the result contained in Eqs.~\eqref{wzor:DOP}
and~\eqref{wzor:defalpha2} is rather involved 
and requires a numerical 
solution for the exciton states, the essential features of our
theoretical predictions are already clear in a simplified,
semi-quantitative treatment\cite{kaczmarkiewicz11a}. 
If the Coulomb coupling is neglected then for a single bright
exciton state with both the electron and the hole 
in the single particle state $i_{0}$ one has
$c^{(\beta)}_{(i,3/2)(j,1/2)}=\delta_{ii_{0}}\delta_{ji_{0}}$. In this case, if the electron and hole
envelope wave functions are assumed identical  then 
$\alpha^{(\beta)}_{3/2,1/2}=1$ and 
\begin{displaymath}
\im \alpha^{(\beta,\mathrm{k})}_{-1/2,1/2}=
 \frac{\sqrt{3}\hbar^2}{2 m_0 \Delta E_{\mathrm{lh}}} \gamma_3
\left(
\langle k_l^2\rangle_{i_{0}}-\langle  k_t^2\rangle_{i_{0}}\right),
\end{displaymath}
where $\langle \ldots\rangle_{i_{0}}$ denotes the average value in the
envelope state $i_{0}$.
 Typically (assuming approximately box-like confinement model), 
\begin{displaymath}
\langle k_l^2\rangle_{i_{0}}\sim 
\pi^{2}\frac{n_{\mathrm{l}}^{2}}{L^{2}}\quad
\langle  k_t^2\rangle_{i_{0}}\sim 
\pi^{2}\frac{n_{\mathrm{t}}^{2}}{D^{2}},
\end{displaymath}
where $L$ and $D$ are the confinement dimensions
along and perpendicular to the QDash, respectively and
$n_{\mathrm{l}},n_{\mathrm{t}}$ are the quantum numbers associated
with the excitations along and perpendicular to the QDash elongation.
Therefore, from Eq.~\eqref{wzor:DOP}, one
can estimate the kinetic contribution to the DOP of light emitted from
a single state in this case 
\begin{equation}\label{DOP-estim}
\Pi_{\mathrm{k}}(n_{\mathrm{l}},n_{\mathrm{t}})=
\frac{\hbar^{2}\pi^{2}}{m_{0}\Delta E_{\mathrm{lh}}}\gamma_{3}
\left( \frac{n_{\tra}^{2}}{D^{2}}
-\frac{n_{\lon}^{2}}{L^{2}} \right).
\end{equation}
Of course, for other confinement models, the quantum numbers will
enter in Eq.~\eqref{DOP-estim} in a different way (e.g., linear for a
harmonic oscillator). The main conclusions will, nonetheless, be
the same.

The approximate formula \eqref{DOP-estim} captures the essential effect of the
confinement shape on the polarization properties of the emitted
light. For an in-plane isotropic confinement ($L=D$), there is no
preferred polarization axis. For an anisotropic structure, the light
is preferentially polarized along the structure, to the degree
dependent on the occupation of the higher excited states. It turns out
that this simple model is sufficient to account for the
high-temperature spectrally resolved degree of polarization for an
ensemble of QDashes \cite{kaczmarkiewicz11a}. However, for the study
of more subtle effects, like additional localization, one needs a more
detailed approach, including a realistic confinement shape, the
Coulomb correlations, and the thermal distribution of occupations. 
Therefore, in the following section we introduce a more detailed and
accurate model of electron and hole confinement as well as
Coulomb-coupled exciton states. While Eq.~\eqref{DOP-estim} will be
useful for the qualitative interpretation of the results, the general
Eq.~\eqref{wzor:DOP} with the wave
functions derived in Sec.~\ref{sec:theory-model} will be used in the
numerical modeling of the polarized emission that will be compared
with the experimental results.

\subsection{Model of a QDash}
\label{sec:theory-model}

In order to obtain quantitative estimates of the degree of
polarization of the radiation emitted by a QDash structure occupied by
a single exciton we now introduce a simple model of the system. This will
allow us to calculate the kinetic contribution to the polarization,
which critically depends on the character of the wave functions. The
strain-related part $\Pi_{\mathrm{s}}$ is assumed to be similar for all
the states as discussed above. Therefore, this part will be treated as
a constant parameter to be found from fitting to experimental
results. 

Our choice here is to model the carrier trapping by width
fluctuations, which are the dominating type of confinement
irregularity in some InAs/InP QDash samples \cite{ooi08}. In the
Appendix, we 
show that localization appears also on the zigzag corners.  In fact,
the exact localization mechanism is not essential for our
conclusions. Our choice is convenient for our present purpose as the
localization on geometrical fluctuations can be consistently modeled
within a simple approach, while trapping on the bends may involve
nontrivial strain fields and, therefore, require much more detailed
structural modeling, which is beyond the scope of this paper.

\begin{figure}[!tb]
\begin{center}
\includegraphics[width=85mm]{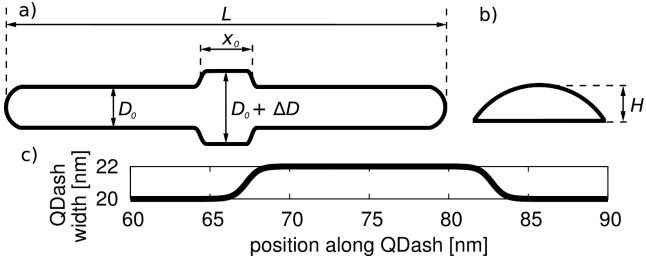}
\end{center}
\caption{\label{fig:diagram} (a,b) A schematic representation of the QDash
  geometry used in our model: top (a) and section (b) view. (c) An
  exact plot of the QDash width as a function of position along its
  length (only the central section of the structure is shown).} 
\end{figure}

The Hamiltonian for a single carrier (electron or hole) confined in a QDash in the
single-band effective mass and envelope function approximations is
\begin{displaymath}
H_{\alpha}=-\frac{\hbar^2}{2m^*_{\alpha}} \Delta + V(\bm{r}),
\end{displaymath}
where $\alpha$ denotes the type of a carrier
($\alpha=\mathrm{e,h}$). The QDash confinement 
potential $V(\bm{r})$ is modeled as a 3D potential well with the shape
reproducing the essential features of the QDash geometry, in
particular the presence of a widening that can trap the carriers (see
Fig.~\ref{fig:diagram}). 
We assume that the QDash structures have the cross-section in the
form of a circular segment, with the base width (chord length)
changing along 
the QDash length (the $x$ coordinate) according to 
\begin{displaymath}
D(x)= D_{0}+\frac{\Delta D (1+4e^{-b})}{1+4e^{-b}\cosh(2bx/x_{0})}, 
\end{displaymath}
where $D_{0}$ is the QDash base width away from the widening,
$\Delta D$ is the depth of the fluctuation,
$x_{0}$ is the length of the fluctuation, and $b$ defines the
shape of the widening (we choose $b=20$; see Fig.~\ref{fig:diagram}(c)
for the shape of $D(x)$).
In our model, the widening of the structure is located symmetrically
in the center of the QDash
(preserving the $D_2$ symmetry of the structure). A non-central location
of the widening modifies the details of the spectrum due to different
selection rules but does not lead to essential modifications of the
polarization properties discussed here.
The height of the dash is $H(x)=\zeta D(x)$, where 
$\zeta$ is a constant height to width ratio.
For the numerical computations we use standard material parameters for
the InAs/InP material pair\cite{lawaetz71} and parameter values
computed for strained nanostructures in this material system
\cite{holm02}:
The effective band offsets between the materials (including strain
effects), defining the depth of the 
confinement potentials, are
taken as 400~meV and 250~meV for electrons and holes,
respectively. The effective masses used in our calculations are $0.037m_0$
for electrons and $0.33m_0$ for holes, where $m_{0}$ is
the free electron mass. The Luttinger parameter is $\gamma_{3}=9.29$. 

The approximate single-particle envelope wave functions for electrons
and holes are found using a variational 
method generalizing the ``adiabatic'' approximation \cite{wojs96}, using
the fact that the confinement along the $x$ direction is much
weaker than in the other two directions and changes smoothly. First,
for each $x$, we 
variationally minimize the single-particle Hamiltonian (where we
suppress the carrier type index $\alpha$ for clarity)
\begin{displaymath}
H_{yz}= -\frac{\hbar^2}{2m^*} 
\left( \frac{\partial^{2}}{\partial y^{2}}
  +\frac{\partial^{2}}{\partial z^{2}} \right)  + V(\bm{r})
\end{displaymath}
in the class of 2-dimensional harmonic oscillator (2DHO) ground state
wave functions 
\begin{eqnarray}
\lefteqn{\psi_{0}(y,z;x)=}\nonumber \\
&&\frac{1}{\sqrt{l_z(x) l_y(x) \pi }} \exp 
\left\{ -\frac{[z-z_0(x)]^2}{2l^{2}_z(x)}
-\frac{y^2}{2l^{2}_y(x)} \right\},
\label{wzor:psiZ}
\end{eqnarray}
with the variational parameters $l_{y}(x)$ and $l_{z}(x)$
corresponding to the wave function widths in  the
$y$ and $z$ direction and $z_{0}(x)$ representing the position of the center of
the wave function along $z$. 
Next, we generate the set of $x$-dependent effective potentials
\begin{equation}\label{wzor:EffPot}
\epsilon_n(x)=  \int dz \int dy
\psi_{n}^{*}(y,z;x)H_{yz}\psi_{n}(y,z;x),
\end{equation}
where $\psi_{n}(y,z;x)$ is the wave function of the 2DHO (with the same
parameters as those obtained variationally for the ground state)
representing the $n$th state along $y$. 
This allows us
to approximately account for the excited states along $y$ (however, we 
restrict the dynamics in the 
strongest confinement direction $z$ to the ground state). The
approximate energies $\epsilon_{n}(x)$ are 
 then treated as effective potentials for the one-dimensional
 eigenvalue equations in the QDash elongation direction,
\begin{equation}\label{wzor:RnieEfektywn}
\left [ -\frac{\hbar^{2}}{2m^*}\frac{\partial^{2}}{\partial x^{2}} 
+ \epsilon_n(x) \right ] f_{nm}(x) = E_{nm} f_{nm}(x),
\end{equation}
which are solved numerically (separately for each $n$).
The full envelope wave functions are then
\begin{equation}\label{wzor:wavefunctions}
\psi_{nm}(x,y,z)=\psi_{n}(y,z;x)f_{nm}(x).
\end{equation}
In the following, both quantum numbers $n$ and $m$ will be denoted by a
single index $i$. 

\begin{figure}[!tb]
\begin{center}
\includegraphics[width=70mm]{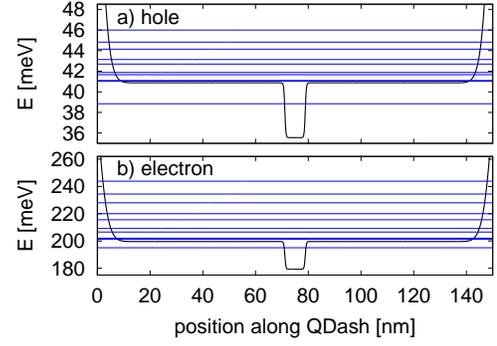}
\end{center}
\caption{\label{fig:levels}(Color online)
The one-dimensional effective potential along the QDash
$\epsilon_{0}(x)$ and the spectrum of single-particle levels for the hole (a) and
electron (b).}  
\end{figure}

In Fig.~\ref{fig:levels}, we show the one-dimensional effective
potential along the QDash $\epsilon_{0}(x)$, corresponding to the
lowest states in the $yz$ plane, and the spectra of
single-particle electron and hole levels. As can be seen, for the
parameter values chosen here, there is one trapped state both for the
electron and for the hole.

Based on the calculated single-carrier states, one can construct the
product basis for the excitonic states and diagonalize the system
described by the Hamiltonian 
\begin{eqnarray}
H &=& \sum_{i} E_{i}^{(e)} a_{i}^\dag a_{i}+\sum_{i}
E_{i}^{(h)} h_{i}^\dag h_{i } \nonumber \\ 
  & &+ \sum_{ijkl} V_{ijkl} a_{i}^\dag h_{j}^\dag h_{k} a_{l},
\label{wzor:ham_ex}
\end{eqnarray}
where 
$a_{i}^{\dag},a_{i}$ and $h_{i}^{\dag},h_{i}$ are electron and hole
creation and annihilation operators, respectively,
$E_{i}^{e,h}$ are the energies found from
Eq.~\eqref{wzor:RnieEfektywn} for electrons and holes,  
and $V_{ijkl}$ are the matrix elements of the electron-hole
interaction, 
\begin{eqnarray*}
V^{e-h}_{ijkl} &=& \langle ij | H_{\mathrm{e-h}} | kl \rangle \\
&=& - \int d^3 r_e \int d^3r_h 
\psi_{\mathrm{v}i}^{*} (\bm{r_{\mathrm{e}}}) 
\psi_{\mathrm{c}j}^{*} (\bm{r_{\mathrm{h}}})\\
&& \times \frac{e^2}{4 \pi \varepsilon \varepsilon_0}
\frac{1}{|\bm{r_e}-\bm{r_h}|}  
\psi_{\mathrm{c}k}(\bm{r_{\mathrm{h}}}) 
\psi_{\mathrm{v}l}(\bm{r_{\mathrm{e}}}),
\end{eqnarray*}
where $\varepsilon_{0}$ is the vacuum permittivity and $\varepsilon$
is the relative dielectric constant of the QDash material
($\varepsilon=14.6$ for InAs). 
Upon the diagonalization of the 
Hamiltonian given by Eq.~(\ref{wzor:ham_ex}) one obtains the 
coefficients $c_{(i,3/2)(j,1/2)}^{(\beta)}$ defining the exciton eigenstates which
are then used in the calculation of the DOP according to
Eq.~\eqref{wzor:DOP} (in the leading order in subband mixing). 

The inhomogeneous ensemble of QDashes is modeled by assuming a
Gaussian distribution of the sizes (corresponding to the formal
distribution function $f(\eta)$ introduced in
Sec.~\ref{sec:theory-mixing}) with the standard deviation of 10\%
and with a proportional scaling of all the dimensions. The average
length of a QDash is taken to be $L=150$~nm, the QDash lateral aspect
ratio is $L/D_{0}=6$, the height to width ratio is $\zeta=1/5.5$, and
the size of the width fluctuation is $x_{0}=8$~nm, which corresponds
to a typical geometry of real structures \cite{reithmaier07,ooi08}.
In our numerical modeling, $\Delta D/D_{0}$ and
$\Delta E_{\mathrm{lh}}$ are adjustable parameters. 

For the numerical calculations, we choose a set of single carrier
states with lowest eigenenergies. In order to achieve convergence 
we restrict our single carrier states to about 25 electronic states 
(with the dynamics along the transverse direction restricted to the
ground state),  
and 50 hole states. Since the effective mass of heavy holes is larger than
that of an electron, also states with excitations along $y$ direction
($n=2$), which are coupled to the $n=0$ states by the Coulomb
interaction, are included. Depending on the exact values of the QDash 
shape parameters, the number of states included in the calculations may
slightly vary. Luminescence at finite temperatures is calculated 
using Boltzmann distribution for the occupation of exciton
states. This is equivalent to the assumption that thermalization is
fast enough to assure full relaxation to equilibrium at the lattice
temperature on time scales much shorter than the exciton life
time. This should be the case in the system under discussion since the
spectrum of a strongly elongated structure is rather dense and the
optical response originates from structures where electron-hole pairs
are captured, which enables the efficient electron-hole
scattering-assisted relaxation channels. 

\begin{figure}[!tb]
\begin{center}
\includegraphics[width=80mm]{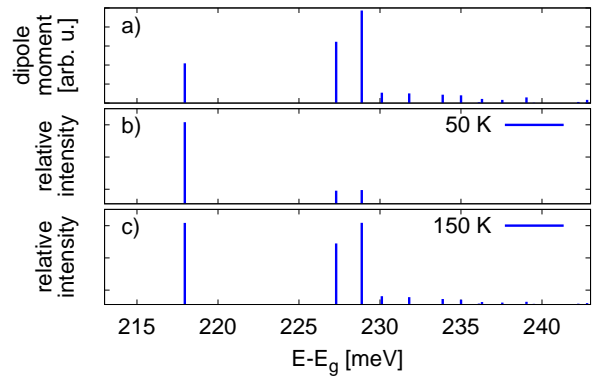}
\end{center}
\caption{\label{fig:dipole-moments}(Color online)
(a) Dipole moments for optical transitions for the exciton states
obtained from numerical calculations for a 150~nm long QDash with
$D/D_{0}=0.1$. (b) and (c) Emission spectrum from a single QDash
taking into account thermal distribution of the carriers at two
different temperatures. Here, relative intensities (normalized to the
intensity of the ground state transition) are shown.}   
\end{figure}

In Fig.~\ref{fig:dipole-moments}, we show the optical properties of a
single QDash obtained from our numerical
calculations. Fig.~\ref{fig:dipole-moments}(a) presents the interband dipole
moments (oscillator strengths) for various exciton states confined in the
QDash. In Fig. ~\ref{fig:dipole-moments}(b,c), the computed emission
spectrum at finite temperatures is shown. The lowest energy line
corresponds to a state trapped in the width fluctuation while the
higher energy lines are related to states delocalized in the whole
QDash volume. The lowest delocalized states have larger oscillator
strengths than the trapped ground state but because the latter is
down-shifted by the trapping and Coulomb energy the contribution from
the delocalized states becomes large only at relatively high
temperatures. Note, however, that the appearance of the higher energy
emission lines in a single QDash spectrum some 10~meV above the
fundamental transition does not considerably affect the ensemble
emission shown in Fig.~\ref{fig:eksp-PL} because of the large
inhomogeneous broadening of about 50~meV. As a result, the ensemble
emission feature is rather symmetric at all temperatures, reflecting
the symmetric ensemble distribution of QDash morphology features,
which is common for this material system \cite{sauerwald05}.

\section{Results and discussion}
\label{sec:results}

In this section we present the results of measurements and theoretical
modeling that aim at a complete characterization and explanation of
the polarization properties of the luminescence from QDashes,
as a function of temperature and excitation power. 
It should be noted that our theory presented in Sec.~\ref{sec:theory}
is formally limited to the low power regime, where a single exciton
model for a QDash emission is valid. 
Taking into account a reasonable estimate of losses, the power density
and laser spot size as given in Sec.~\ref{sec:exp} corresponds
to the photon fluence rate on the order of 10$^{6}$/(cm$^{2}\cdot$ns).
With the average exciton lifetime of 1~ns, one gets an estimate of
the QDash occupation on the order of $10^{-5}$ (assuming the exciton 
lifetime of 1~ns) for the lowest excitation power used. Double
occupation under weak excitation conditions is therefore highly
improbable, even if one allows for the extension of charge life time
due to dark spin configurations and trapping of unpaired
carriers. Hence, the 
single-exciton theory provides a basis for the interpretation
of the temperature dependence of the DOP at low powers. However,
qualitative or 
semi-quantitative interpretation of power-dependent spectra is also
possible, based on the concept of state filling (due to Pauli
blocking) and on the general tendencies predicted by
Eq.~\eqref{DOP-estim}.  

\begin{figure}[!tb]
\begin{center}
\includegraphics[width=80mm]{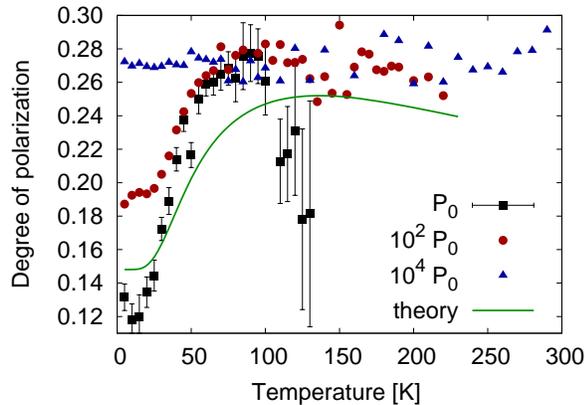}
\end{center}
\caption{\label{fig:eksp-temp}(Color online)
Points: Temperature dependence of emission
  degree of polarization from the ensemble of InAs/InP quantum
  dashes in three different excitation regimes as indicated. For
  clarity, the error bars are shown only for the lowest power data
  where the measurement uncertainty is the largest. Solid line: the
  theoretical result.}  
\end{figure}

In order to systematically study the dependence of the DOP on the
excitation conditions and temperature we 
have measured the DOP as a function of temperature 
for three different excitation powers, shown in
Fig.~\ref{fig:eksp-temp}. When 
a very low excitation power is
used the DOP at low temperatures is on the level of 0.12-0.13 and then
increases with temperature
up to about 0.27 at above 60~K (which roughly corresponds to 5~meV
activation energy). The dependence is similar
for the intermediate excitation power 
but the initial (low
temperature) value is higher (about 0.19). Eventually, the DOP is almost
constant on the level of 0.27 over the entire temperature range for a
very high excitation.

These results are consistent with the emission properties from either
locally trapped or delocalized (in the whole QDash volume)
states.
Indeed, as discussed in Sec.~\ref{sec:theory}, the DOP for the
emission from a given state depends on the
imaginary part of the relevant matrix element of the
$\bm{k}\cdot\bm{p}$ Hamiltonian given by Eq.~\eqref{wzor:kpim} which,
in turn, is determined by the values of the wave vector components
$k_{\mathrm{l}}$,$k_{\mathrm{t}}$ (along and perpendicular to the QDash
elongation) characteristic for the relevant wave function. 
At low temperature and low excitation, the detected emission originates
mainly from the ground state excitons localized at the QDash potential
fluctuations, which are more isotropic than the dash itself. Due to
the reduced anisotropy, $\kl\sim \kt$, hence the kinetic component to the
DOP is small. This results in a small DOP, 
similar to that observed for slightly asymmetric quantum dots \cite{cortez01}.
By increasing the temperature, the excitons are thermally released into
the whole QDash volume. For such low-energy extended states the confinement
along the QDash is much weaker than across the structure. Since
$\kl\sim 1/L$ and $\kt\sim 1/D_{0}$ these two wave vector components
become imbalanced and hence the emission becomes more linearly
polarized. This effect
is also clear from the simplified 
Eq.~\eqref{DOP-estim} if one substitutes for $L$ the effective
confinement size for a given state ($L\sim D$ for a trapped ground
state, $L\gg D$ for a delocalized higher energy state).

With increasing excitation power, the higher states with anisotropic
wave functions become filled and
contribute to luminescence already at low temperatures. Therefore, 
the low temperature DOP increases and the amplitude of the
temperature-dependent change is reduced until, 
for the highest excitation, the contribution of the
localized excitons to the emission becomes negligible and the DOP does
not significantly change with temperature. Remarkably, the saturation
level for the DOP is the same for all the excitation powers, i.e., it
is insensitive to the excitation conditions, reflecting the intrinsic
asymmetry of the carrier states trapped in the QDash.

The results obtained from the full theoretical model presented in
Sec.~\ref{sec:theory} using the exciton wave functions found in
Sec.~\ref{sec:theory-model} and the general equations
\eqref{wzor:defalpha2} and \eqref{wzor:DOP} for a certain set of
parameters are shown by 
the green solid line in Fig.~\ref{fig:eksp-temp}. As our theory is
restricted to single-exciton states the theoretical
curve corresponds to the low excitation limit.
The S-shaped temperature dependence is reproduced if one
assumes a QDash ensemble with an appropriately chosen relative amplitude
of the widening. For the curve in Fig.~\ref{fig:eksp-temp},
$\Delta D/D_{0}=0.1$. 
For this value of the shape parameter, only the exciton ground state
exhibits highly localized properties, which are reflected by a
moderate value of the calculated DOP for low temperatures. 
Clearly, the model is able to reproduce all
the qualitative features of the measured dependence: 
When only the ground state is occupied
the degree of polarization of emitted radiation is small. At higher
temperatures, emission from higher energy excitonic states
(extended over the whole area of a QDash) contributes, leading to
the increased value of DOP. 
The theoretical result shown in
Fig.~\ref{fig:eksp-temp}, is obtained for 
the separation between light end heavy holes $\Delta
E_{\mathrm{lh}}=20$~meV (the amplitude of the DOP change scales
inversely proportional to this parameter). While this value is rather
small, one should 
notice that this energy is an effective parameter averaged over all
the heavy hole states contributing to the luminescence and should not
be treated as the separation between the lowest heavy hole state and
the light hole subband (in the decoupled band picture). 

As can be seen, our model is also
able to account to some extent for the observed decrease of the DOP at high
temperatures for lower excitation powers.
This effect can again be understood with the help of
the simplified Eq.~\eqref{DOP-estim}.
As the temperature grows, the first process is the release of carriers
from the additional confinement center which leads to an increase of
the effective confinement length $L$ in Eq.~\eqref{DOP-estim} as
discussed above. At first, the lowest lying states are filled so that
$n_{\mathrm{l}}\sim 1$. However, when the temperature further increases
higher excited states are occupied with growing values of
$n_{\mathrm{l}}$, while $n_{\mathrm{t}}$ remains small due to the
stronger confinement and higher excitation energy in the transverse lateral 
direction. It is clear from Eq.~\eqref{DOP-estim} that this increase
of $n_{\mathrm{l}}$ reduces the value of the DOP as indeed observed in
the low-power experimental data and in the theoretical result. 
Quantitatively, the strong decrease of the low-excitation DOP at high
temperatures is not reproduced by the theory probably due to the 
simplified nature of our QDash model which does not properly account
for the states in the higher energy sector. Moreover, one needs to
keep in mind that the signal values at these conditions are very low,
hence the uncertainty of the measured DOP is very high in this range
(see error bars in Fig.~\ref{fig:eksp-temp}).

\begin{figure}[!tb]
\begin{center}
\includegraphics[width=80mm]{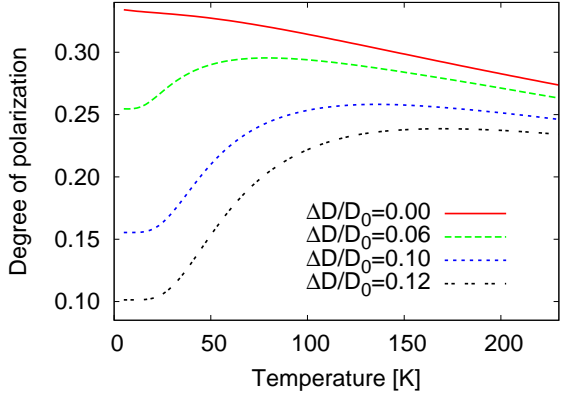}
\end{center}
\caption{\label{fig:cmp}(Color online) Influence of the amplitude of the widening of
  a QDash on the calculated DOP of the system. The strain contribution
is assumed the same in all the cases.} 
\end{figure} 

Let us emphasize that the additional localization is
essential for the explanation of the observed
polarization properties of the PL from QDash structures. The
experimental results cannot be reproduced even qualitatively in the
absence of additional confinement inside a QDash. 
When the shape of a QDash is
more uniform the ground state of the system 
is not trapped and has a rather anisotropic wave function, similar to
all the other states in the structure. As shown in Fig.~\ref{fig:cmp},
this results in 
a relatively high degree of polarization already at low temperatures,
which decreases at higher 
temperatures due to increasing occupation of higher excited states. This
kind of behavior is a fingerprint of a nanostructure with a  
uniform width. When the widening of a QDash is large enough
to confine carriers the results change considerably. Thus, the
character of
polarization properties of emitted radiation reflects the shape of
confining potential: Different confinement conditions lead to
qualitatively different polarization properties of luminescence from
the system. Interestingly, the saturation level and the
high-temperature behavior are different for different sizes of the
shape fluctuation, which shows that the presence of the trapping
potential is important even when the emission from the delocalized states
becomes dominant.

\begin{figure}[!tb]
\begin{center}
\includegraphics[width=80mm]{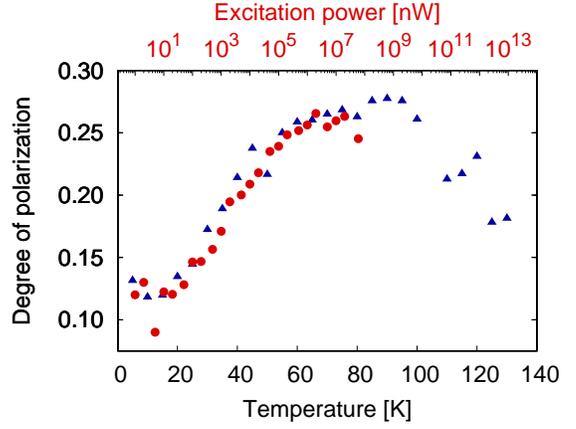}
\end{center}
\caption{\label{fig:eksp-compar}(Color online) The comparison between the temperature
  (bottom horizontal axis, blue triangles) and excitation power (top
  horizontal axis, red circles) dependence of emission degree of
  polarization for the ensemble of InAs/InP quantum dashes.}    
\end{figure}

Effects similar to those observed in the temperature dependence can be
seen in the excitation power dependence of the linear polarization
(red circles in Fig.~\ref{fig:eksp-compar}). The DOP increases and
finally saturates on the same level as in the temperature
dependence. Obviously, the cases of high temperature and high
excitation power are physically different: In the former, we deal with
a single exciton system in which the occupation of higher states is
due to thermal redistribution. In the latter, one has a many-body
system with excited state occupation forced by state filling. The
correspondence between the DOP observed in these two essentially
different cases provides a strong confirmation that the effect is due
to increased occupation of
excited states, which is a common factor of these two physical
situations: Since the higher states have anisotropic wave functions
they contribute 
strongly to polarized emission. A straightforward comparison of
temperature and excitation power influence on the DOP is presented in
Fig.~\ref{fig:eksp-compar} where the DOP axis is common and the
temperature and excitation 
power axes (horizontal ones) are translated with respect to each other
and scaled in order to emphasize the correspondence between the shapes
of the curves.  This figure shows directly that the two external
factors give a similar effect if the excitation power
values are presented in a logarithmic scale. This is consistent with
the fact that the occupations of excited states scale with the
excitation power
according to a power law, while their dependence on temperature is
exponential. Therefore, the agreement between the two dependencies
supports the idea that the distribution of carriers over the confined
states in the QDash is crucial for the polarization properties of the
emitted light: Higher excitation power creates more electron-hole pairs
which occupy higher energy states according to Pauli exclusion principle,
whereas elevated temperatures result in the redistribution of carriers
already existing in the system allowing them to occupy higher energy
levels. Essentially, however, both mechanisms lead to the 
observation of radiative recombination of carriers from higher-energy
states that are expected to be confined in the whole QDash volume.  
At the same time, the logarithmic scaling between the temperature and
power dependence is a strong argument in favor of
state filling as the mechanism of the power dependence, as opposed to
heating. If the excitation power acted via heating the local 
temperature would be proportional to the power and the dependencies on
the power and temperature would scale linearly, which is not the case.  

\section{Conclusions}
\label{sec:concl}

We have investigated polarization-resolved luminescence from an
ensemble of InAs/InP quantum dashes. We have developed a theory
relating the observed degree of linear polarization to the
asymmetry of the carrier wave functions via the anisotropy effect on
the hole subband 
mixing. By comparing the temperature and power dependent degree of
polarization to the theoretical predictions we were able to conclude
that the lowest carrier states usually have a much lower degree of
asymmetry than the structure itself, which can be related to
additional trapping of the excitons to potential fluctuations within
the dash volume. Thus, our findings reveal a nontrivial character of
carrier states in these systems of technological and applicational
relevance. 

Our analysis shows, in addition, that the cw polarization-resolved
spectroscopy can be used as a probe of the localization effects within
such elongated objects. Thus, this widely available optical tool can
yield important information on the actual properties of carrier states
in anisotropic systems.

Apart from the general interest in the electronic structure of quantum
dashes, the 
knowledge on their polarization properties and the 
nature of the carrier states is essential for modeling the emission
properties of such
nanostructures and can be important for the operation of
some optoelectronic devices, especially in those applications where
polarization control is important (e.g., polarization-insensitive
optical amplifiers \cite{saito97}). The strongly confined character of the lowest
carrier states in QDashes will also have a considerable impact on the
performance of futuristic photonic devices which are based on quantum
electrodynamics experiments
\cite{balet07,chauvin09,lamy09,reitzenstein10}. It will affect, 
for instance, the exciton 
fine structure and anisotropy splitting energies, the energies and
even relative spectral position of the excitonic complexes in a QDash,
and finally the kinetics of the transitions between the excitonic
states. The effect of carrier (exciton) trapping will affect both the
total exciton oscillator strength and the polarization selectivity
which become crucial when QDashes are used as quasi-zero-dimensional
emitters placed inside a microcavity, suitable for single photon
sources in the telecommunication wavelength range of 1.3--1.55 $\mu$m.

\begin{acknowledgments}
This work has been supported by the Polish Ministry of Science and
Higher Education within Grant No. N N202 181238, by the COPERNICUS
award of the Foundation
for Polish Science and Deutsche Forschungsgemeinschaft, and by the
State of Bavaria. PK and PM acknowledge
support from the TEAM programme of the Foundation for Polish Science,
co-financed by the European Regional Development Fund. 
\end{acknowledgments}

\appendix

\section{Localization on zig-zag corners}

In this Appendix, we verify that zig-zag bends act as localization
centers in a similar way to local widenings. Thus, the theory
presented in this paper is applicable also to QDash systems where
bends, rather than widenings, are the dominating kind of shape
irregularities. Here, we aim at demonstrating the localization effect
qualitatively, therefore we present a simple model, including only the
geometrical features of the confinement.  It should be kept in mind,
however, that in a real system a specific strain distribution will
occur in the bend area, which must be taken into account in more
realistic modeling.

\begin{figure}[!tb]
\begin{center}
\includegraphics[width=85mm]{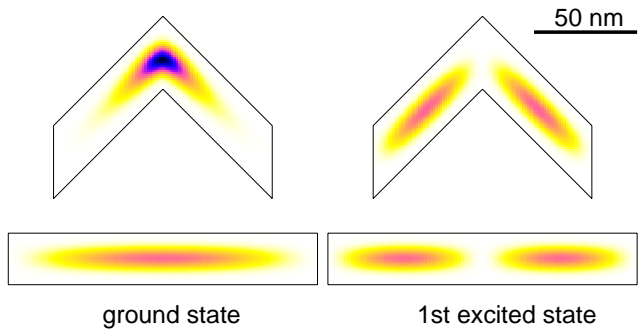}\end{center}
\caption{\label{fig:zigzag}(Color online) The electron densities
  in the $xy$ plane (integrated over $z$) for the ground and first
  excited state of a bent and straight dash. }    
\end{figure}

We consider a QDash of dimensions as in the main body of the paper
with a homogeneous cross section but with a 90$^{\circ}$ bend present
in the middle (see the outlines in the upper part of
Fig.~\ref{fig:zigzag}). The 
electron confinement is modeled as a three-dimensional potential box
with the confinement potential depth and cross-section geometry
as  described in Sec.~\ref{sec:theory-model}. The
electron wave functions are found in a way similar to that presented
in Sec.~\ref{sec:theory-model} (see also Fig.~\ref{fig:diagram}) but
this time only the one-dimensional 
Schr\"odinger equation along the strongest confinement direction $z$
is solved, yielding a two-dimensional effective potential for the eigenvalue
problem in the $xy$ plane. The latter is solved by expansion into
plane waves. 

In Fig.~\ref{fig:zigzag}(a), we show the resulting probability
densities for the ground and first excited electron state for the bent
dash, as compared with a straight, rectangular dash. It is clear
that the ground state gets localized on the corner, while the first
excited state remains delocalized in the same way for both the bent
and straight QDash. 


\end{document}